\documentclass[12pt]{article}
\usepackage{latexsym,epsfig,graphicx,a4wide,amssymb}

{\rm }

\def\be{\begin{equation}}
 \def\ee{\end{equation}}
 \def\bea{\begin{eqnarray}}
 \def\eea{\end{eqnarray}}

\def\2{\frac{1}{2}}
\def\4{\frac{1}{4}}

\catcode`\@=11

\@addtoreset{equation}{section}

\def\@normalsize{\@setsize\normalsize{15pt}\xiipt\@xiipt
\abovedisplayskip 14pt plus3pt minus3pt%
\belowdisplayskip \abovedisplayskip
\abovedisplayshortskip  \z@ plus3pt%
\belowdisplayshortskip  7pt plus3.5pt minus0pt}
\def\small{\@setsize\small{13.6pt}\xipt\@xipt
\abovedisplayskip 13pt plus3pt minus3pt%
\belowdisplayskip \abovedisplayskip
\abovedisplayshortskip  \z@ plus3pt%
\belowdisplayshortskip  7pt plus3.5pt minus0pt
\def\@listi{\parsep 4.5pt plus 2pt minus 1pt
            \itemsep \parsep
            \topsep 9pt plus 3pt minus 3pt}}

\def\underline#1{\relax\ifmmode\@@underline#1\else
        $\@@underline{\hbox{#1}}$\relax\fi}
\@twosidetrue \relax

\catcode`@=12

\catcode`\@=11

\def\section{\@startsection{section}{1}{\z@}{3.5ex plus 1ex minus
   .2ex}{2.3ex plus .2ex}{\large\bf}}

\def\ps@headings{\def\@oddfoot{}\def\@evenfoot{}
\def\@oddhead{\hbox{}\hfill
        \makebox[.5\textwidth]{\raggedright\ignorespaces --\thepage{}--
        \hfill }}
\def\@evenhead{\@oddhead}
\def\subsectionmark##1{\markboth{##1}{}}
}

\ps@headings

\catcode`\@=12

\begin{document}

\begin{titlepage}
%
%


%

\begin{centering}
\vspace{1cm}
{\Large {\bf  Black Hole Solutions in 5D Horava-Lifshitz Gravity }}\\

\vspace{1.5cm}

{\bf George Koutsoumbas $^{\sharp}$}, {\bf Eletherios
Papantonopoulos $^{*}$}, {\bf Pavlos Pasipoularides $^{\ddag}$},
{\bf Minas Tsoukalas $^{\flat}$} \\
 \vspace{.2in}
 Department of Physics, National Technical University of
Athens, \\
Zografou Campus GR 157 73, Athens, Greece \\

\vspace{3mm}

\end{centering}
\vspace{1.5cm}

\begin{abstract}
We study the full spectrum of spherically symmetric solutions in
the five dimensional non-projectable Horava-Lifshitz type gravity
theories. For appropriate ranges of the coupling parameters, we
have found several classes of solutions which are characterized by
an $AdS_5$, $dS_5$ or flat large distance asymptotic behaviour,
plus the standard $1/r^2$ tail of the usual five-dimensional
Schwarzschild black holes. In addition we have found  solutions
with an unconventional short or large distance  behaviour, and for
a special range of the coupling parameters   solutions which
coincide with  black hole solutions of conventional relativistic
five-dimensional  Gauss-Bonnet gravity.
\end{abstract}

\vspace{1.5cm}
\begin{flushleft}

$^{\sharp}~~$ e-mail address: kutsubas@central.ntua.gr\\
$^{*}~~$ e-mail address: lpapa@cetral.ntua.gr \\
$^{\ddag}~~$ e-mail address: paul@central.ntua.gr \\
$^{\flat}~~$ e-mail address: minasts@central.ntua.gr\\

\end{flushleft}
\end{titlepage}

\section{Introduction}

A novel quantum gravity model, which claims power-counting
renormalizability, has been formulated recently by Horava~
\cite{Horava:2008ih}. This scenario is based on an anisotropy
between space and time coordinates, which is expressed via the
scalings $t\rightarrow b^z t$ and  $x\rightarrow b x$, where $z$
is a dynamical critical exponent. For $z\neq1$ the UV behaviour of
the model is governed by a non-standard Lifshitz fixed point,
while for $z=1$ we recover the well-known free Gaussian fixed
point. In the Horava model, for three spatial dimensions, the
suitable choice is $z=3$.

It is worth noting that in the Horava-Lifshitz (HL) gravity the
four-dimensional diffeomorphism  invariance of general relativity
is sacrificed in order to achieve power-counting
renormalizability. The action of the model can be split into a
kinetic plus a potential term, which both respect a restricted
$(3+1)$ diffeomorphism invariance. The interesting feature is that
the kinetic term contains only second order time derivatives,
while the potential term consists of higher order spatial
derivatives of the metric components. This particular structure
improves significantly the UV properties of the graviton
propagator, and renders the model power-counting renormalizable.
Moreover,  in this way we avoid ghost modes which are usual in
conventional higher order gravity models.

For the construction of the potential term, there has been
proposed \cite{Horava:2008ih}
the so called "detailed balance principle", which is inspired
by condensed matter physics.
The main advantage of this approach is the restriction of the
large number of arbitrary couplings that
appear in the bare action of the model. However, the physical motivation
for  a consideration such as the "detailed balance"  is not
clear \cite{Kiritsis:2009sh,Sotiriou:2009bx}.
An alternative way for constructing an action is to include all
possible operators which are
compatible with the renormalizability of the model; this implies
that all operators with dimension
less or equal to six are allowed in the action (for the exact form of the action see \cite{Sotiriou:2009bx}).

As it has been already mentioned, the HL gravity violates local
Lorentz invariance in the UV, however it is expected that general
relativity is recovered in the IR limit. This implies a very
special renormalization group flow for the couplings of the model.
In particular, the parameter $\lambda$ in the kinetic term of the
action (which measures the departure from the Lorentz invariance)
 should flow to unity, while the higher order
couplings should vanish, or they should become appropriately
small, in the IR. Note that, even though phenomenology suggests a
particular IR limit there is no theoretical study supporting this
behaviour.

In addition, there are several other possible inconsistencies in
HL gravity which have been discussed in several works. In
particular, the absence of full diffeomorphism invariance
introduces an additional scalar mode which can lead to strong
coupling problems or instabilities (see for example
\cite{Charmousis:2009tc}-\cite{ Bellorin:2010je}
and references therein). However, these problems will not be
addressed in the present paper.

Apart from these problems, the HL gravity is an interesting
quantum gravity theory, which has stimulated an extended research
on cosmology and black hole solutions
\cite{Kiritsis:2009sh}, \cite{Calcagni:2009ar}-\cite{Myung:2009dc}.
In addition to general relativity studies, quantum field theory
models in flat space-time with anisotropy have also been
considered \cite{Visser:2009fg}-\cite{Koslowski:2009bc}.

Before proceeding it is important to mention that the HL gravity
can be separated into two versions which are known as projectable
and non-projectable. In the projectable version the lapse function
$N^2$ depends only on the time coordinate, while in the
non--projectable version $N^2$ is a function of both the space and
time coordinates. Although in General Relativity the projectable
and non-projectable ansatz for the metric are equivalent, since
they are connected via a diffeomorphism transformation, in the HL
gravity the full diffeomorphism invariance is broken and they lead
to two distinct theories.

Issues connected with broken Lorentz invariance were studied in
the spherically symmetric solutions of 4D HL gravity. In the case
of detailed balance, such spherically symmetric solutions were
found \cite{Lu:2009em}, but they exhibited an unconventional large
distance asymptotic behaviour. The correct Schwarzchild-flat
asymptotic behaviour can be recovered if the detailed balance
action is modified in the IR by a term proportional to Ricci
scalar, and the cosmological constant term is considered to be
zero \cite{Kehagias:2009is}. A similar study,  in the case of
non-vanishing cosmological constant, has also been carried out
\cite{Park:2009zra}. A generalization to topological black holes
was obtained in  \cite{Cai:2009pe}. Finally, a systematic study of
static spherically symmetric solutions of 4D HL gravity was
presented in \cite{Kiritsis:2009rx} where the most general
spherically symmetric solution  for $\lambda\neq 1$ and general
coupling parameters was obtained.

In this work we present a full study of spherically symmetric
solutions in the non-projectable version of the five-dimensional
Horava-Lifshitz gravity, for $z=4$. For the construction of the 5D
action we do not use the "detailed balance principle", but we
include all the terms which are compatible with the
renormalizability of the model.  In particular, we can include all
spatial curvature terms with dimension less than or equal to
eight. However, the large number of possible terms, which are
allowed in the action, leads to equation of motion of great
complexity. For this reason we restrict our study only to terms of
up to second order in the curvature. Also, we suppose that in the
IR limit 5D the HL gravity reduces to the 5D General Relativity
plus a bulk cosmological constant. A class of spherically
symmetric solutions of the 5D HL gravity  has been considered
previously \cite{Cai:2009ar}, but only for a very specific choice
of the couplings.

Our main motivation in considering static solutions in 5D HL
gravity is to investigate whether the rich spectrum of black hole
solutions found in 4D (see Ref. \cite{Kiritsis:2009rx})
also persists in 5D. It seems that the
known static solutions of the HL gravity in 4D with $\lambda \neq
1$ do not have any obvious relation with the corresponding static
solutions of the relativistic 4D gravity. In 5D however we found
that there is a class of spherically symmetric solutions which
after a proper identification of coupling parameters coincide with
the known black hole solutions of conventional relativistic 5D
Gauss-Bonnet gravity.

Static solutions of 5D  Gauss-Bonnet theory are well known
\cite{gbpapers}. Among them there is a black hole solution which
has  two branches (for a review see \cite{Charmousis:2008kc}). The
first branch is referred to as the Einstein branch while the
second as the Gauss-Bonnet branch. Both branches coincide in the
Chern-Simons limit. As we will discuss in the following, we find
both branches of solutions and in addition these solutions can
also be obtained for a different combination of coupling
parameters of the quadratic curvature terms than the usual
combination that appears in the relativistic Gauss-Bonnet theory.
We also find the black hole solution corresponding to the
Chern-Simons limit with a particular choice of coupling
parameters. This solution has also been found in \cite{Cai:2009ar}
using the "detailed balance principle".

The paper is organized as follows. In section 2 we write down the
action of 5D HL gravity. In section 3 we derive the equations of
motion. In section 4 we analyze the static spherically symmetric
solutions for a special choice of coupling parameters. In section
5 we study the most general static spherically symmetric solutions
of 5D Horava-Lifshitz gravity and finally section 6 contains our
conclusions.

\section{5D Horava-Lifshitz gravity models}

In this section we introduce the notation for the so-called Horava
gravity models in the case of four spatial dimensions ($d=4$).
These models are characterized by an anisotropy between space and
time dimensions
\begin{equation}
[t]=-z, \quad [x]=-1~,
\end{equation}
where $z$ is an integer dynamical exponent. In order to derive the
action of the model, it is useful to express the space-time metric
in the ADM form
\begin{equation} \label{ADM}
ds^2=-  c^2 N^2 dt^2+g_{ij} \left(dx^i-N^i dt\right)\left(dx^j-N^j
dt\right)~,
\end{equation}
where $c$ is the
velocity of light, with dimension $[c]=z-1$, and spatial
components $dx^i/dt$ $(i=1,2,3,4)$. In addition, $N$ and $N_i$ are
the "lapse" and "shift" functions which are used in general
relativity in order to split the space-time dimensions, and
$g_{ij}$ is the spatial metric of signature (+,+,+,+).  For the
dimensions of "lapse" and "shift" functions we obtain
\begin{equation}
[N]=0, ~~[N_i]=z-1~.
\end{equation} In this paper the dynamical exponent $z$ is set
equal to $4.$ The 5D action of the model is constructed from a
kinetic plus a potential term according to the equation \be
\label{action}
 S=\frac{1}{16 \pi G_5 c }\int dt d^d x \sqrt{|g|} N \left \{ {\cal L}_{K}+{\cal L}_{V}\right\}
\ee in which $d$ ($D=d+1=5$) is the spatial dimension and $G_5$ is
the five dimensional Newton constant.

We stress that the main motivation for considering models of this
type is the construction of a power-counting renormalizable
gravity model. However,  in order to achieve renormalizibility,
and simultaneously keep the time derivatives up to second order,
we have to sacrifice the standard 5D diffeomorphism invariance of
general relativity, which is now restricted to the transformation
\be \label{diffeo} \tilde{x}^{i}=\tilde{x}(x^j,t), ~~
\tilde{t}=\tilde{t}(t)~. \ee The kinetic part in the above
Lagrangian of Eq. (\ref{action}) can be expressed via the
extrinsic curvature as: \be \label{curvature1}
 {\cal L}_{K}= (K^{ij}K_{ij}-\lambda K^2), \quad K_{ij}=\frac{1}{2 N} \left\{-\partial_{t}g_{ij}+\nabla_i N_j+\nabla_j N_i\right\},~~ i,j=1,2,3,4
\ee which is invariant under the transformations of Eq.
(\ref{diffeo}). For the construction of the potential term we will
not follow the standard detailed balance principle, but we will
use the more general approach \cite{Kiritsis:2009sh, Sotiriou:2009bx}, according to
which the potential term is constructed by including all possible
renormalizable operators \footnote{We have ignored the terms which
violate parity, see also \cite{Sotiriou:2009bx}.}, that have dimension smaller or equal to eight,
hence we write \be
 {\cal L}_{V}= {\cal L}_R+{\cal L}_{R^2}+{\cal L}_{R^3}+{\cal L}_{\Delta R^2}+{\cal L}_{R^4}+{\cal L}_{\Delta R^3}+{\cal
 L}_{\Delta^2R^2}~.
\ee where the symbol $\Delta$ is defined as $\Delta=\partial_i\partial_i$ ($i=1,2,3,4$).

The dimensions of the various terms in the Lagrangian read
\be [R]=2,~ [{R^2}]=4, ~[{R^3}]=[{\Delta R^2}]=6, [{R^4}]=[{\Delta
R^3}]=[{\Delta^2R^2}]=8~. \ee In this work we are mainly
interested in the lowest order operator ${\cal L}_R$ and the
operator ${\cal L}_{R^2}$, which contains contributions of second
order in the curvature:
\be
 {\cal L}_R=\eta_{0a}+\eta_{1a} R, \quad {\cal L}_{R^2}
 =\eta_{2a} R^2+\eta_{2b} R^{ij}R_{ij}+\eta_{2c}
 R^{ijkl}R_{ijkl}~,
\ee where we have used the notation $R$, $R_{ij}$ and $R_{ijkl}$
for the Ricci scalar,  the Ricci and the Riemann tensors
($i,j=1,2,3,4$), which correspond to the spatial 4D metric
$g_{ij}$. Note that in the case of three spatial dimensions the
term $ R^{ijkl}R_{ijkl}$ is absent, as the Weyl tensor in three
dimensions automatically vanishes. However, in four spatial
dimensions this term cannot be omitted from the action.

The first term ${\cal L}_R$ is necessary in order to recover 5D
general relativity with a cosmological constant in the IR limit.
The second term ${\cal L}_{R^2}$, includes all possible quadratic
terms in curvature, and becomes important in the short distance
regime of the theory. Moreover, $\eta_{0a}$ plays the role of the
cosmological constant, while $\eta_{1a}$, $\eta_{2a}$,
$\eta_{2b}$, and $\eta_{2c}$ are dimensionful coupling constants
with dimensions \be [\eta_{1a}]=6, ~~
[\eta_{2a}]=[\eta_{2b}]=[\eta_{2c}]=4~. \ee In the present
analysis we ignore higher order Lagrangian terms, of dimension six
and eight. Although, we have not derived the detailed expression
for the Lagrangian, it is worth writing some of the higher order
curvature terms here, \bea
 &&{\cal L}_{R^3}=R^3+R_{ij}R^{ij}R+...,~~{\cal L}_{\Delta R^2}
 =R\triangle R+R^{ij}\Delta R_{ij}+...,\\
  &&{\cal L}_{R^4}=R^4+(R_{ij}R^{ij})^2+...,~~{\cal L}_{ \Delta R^3}
  =R^2\triangle R+...,~~{\cal L}_{\Delta^2R^2}=R \Delta^2 R+...
\eea where $\Delta=\nabla_i\nabla^i$. The short distance effects
of these terms may be important, but this topic is left for a
future investigation.

If this model is to make sense, it is necessary that the 5D
general relativity  (with a cosmological constant in our case) is
recovered in the IR limit. Although there is no theoretical proof
for this difficult question, we will assume that the
renormalization group flow towards the IR leads the parameter
$\lambda$ to the value one ($\lambda=1$), hence 5D general
relativity is recovered. Also, to obtain the Einstein-Hilbert
action \be S_{EH}= \frac{1}{16\pi G_5} \int dx^0 d^4 x \sqrt{g} N
\left (\tilde{K}_{ij}\tilde{K}^{ij}- \tilde{K}^2 + R+\eta_{0a}
\right)~, \ee we have to set $\eta_{1a}=c^2$, and \be
\tilde{K}_{ij}=\frac{1}{2 N} \left\{-\partial_{0}g_{ij}+\nabla_i
\left(\frac{ N_j}{c}\right)+\nabla_j\left(\frac{
N_i}{c}\right)\right\}~. \ee where the time-like coordinate
$x_{0}$ is defined as $x_{0}=c t$.

\section{Equations of motion}

We are looking for 5D spherically symmetric solutions of the
Horava-type gravity model we constructed in the previous section.
We use the following ansatz \footnote{There is a more general
ansatz for the metric, of the form $ds^2=-N(r)^2
dt^2+f^{-1}(r)~(dr+N^{r}(r)dt)^2 +r^2 d\Sigma_{k}^2$ with nonzero
shift $N^{r}(r)$, but we have set $N^{r}(r)=0$
\cite{Kiritsis:2009rx} to simplify the equations.} for the metric
\be \label{metric} ds^2=-N(r)^2 dt^2+f^{-1}(r)~dr^2 +r^2
d\Omega_{k}^2~, \ee in which $r$ is a radius coordinate that
corresponds to the extra dimension, and
$d\Omega_{k}^2$ is the metric of a 3D maximally symmetric space,
where $k$ is the spatial curvature of 3D hypersurfaces  and for
$k=1,-1,0$ we have a sphere, hyperboloid or 3D torus topology
correspondingly. In what follows it is convenient to perform the
transformation \be f(r)=k+r^2 Z(r) \label{Z}~. \ee Then the action
of the model to second order in curvature terms is \be
\label{actionsec}
 S=\frac{1}{16 \pi G_5 }\int dt d^d x \sqrt{|g|} N \left(K^{ij}K_{ij}
 -\lambda K^2+\eta_{0a}+\eta_{1a} R+\eta_{2a} R^2+\eta_{2b} R^{ij}R_{ij}+\eta_{2c} R^{ijkl}R_{ijkl} \right)
\ee which can be put into the form \be \label{actionr}
S\left[N(r),Z(r),\frac{dZ(r)}{dr}\right]=\int_{0}^{+\infty} dr~
L\left[N(r),Z(r),\frac{dZ(r)}{dr}\right]~, \ee after we integrate
out the angular coordinates, where \be
L\left[N,Z,\frac{dZ}{dr}\right]\sim r^3\sqrt{\frac{N^2}{f}}
\left(P\left(r \frac{d Z}{dr}\right)^2+  M(Z) \left(r \frac{d
Z}{dr}\right)+  Q(Z)\right) \ee and the coefficients $P, \ M(Z)$
and $Q(Z)$ are defined by the equations
\begin{eqnarray}
&&P= 3 (3 \eta_{2a} +  \eta_{2b}+ \eta_{2c})~, \nonumber\\
&&M(Z)=6 (12\eta_{2a}+3\eta_{2b}+2\eta_{2c})Z-3\eta_{1a}~,
\nonumber\\ &&Q(Z)=12(12\eta_{2a}+3\eta_{2b}+2\eta_{2c})
Z^2-12\eta_{1a}Z+\eta_{0a}~.
\end{eqnarray}
If we set
\be
\eta=(3 \eta_{2a} +  \eta_{2b}+ \eta_{2c})~, \ \
\varrho=(12\eta_{2a}+3\eta_{2b}+2\eta_{2c})~, \ee we can reduce
the number of the free parameters of the model. This is possible
because of the spherical ansatz for the metric, as it is given by
Eq. (\ref{metric}). In we set $\eta_{1a}=1$, by choosing a
coordinate system in which $c=1$, we obtain the simplified
expressions
\begin{eqnarray} \label{PMQ}
&&P= 3 \eta~, \nonumber\\
&&M(Z)=6\varrho Z-3~, \nonumber \\
&&Q(Z)=12\varrho Z^2-12 Z+\eta_{0a}~.
\end{eqnarray}
We note that the coefficient $P$ is independent from $r$, while
the functions $M(Z)$ and $Q(Z)$ do not depend explicitly on the
radius $r$. The Euler-Lagrange equations for the action
(\ref{actionr}) are \bea \label{lagrange}
\frac{d}{dr}\left(\frac{\partial L}{\partial N'}\right)
-\frac{\partial L}{\partial N}=0~, ~~~
\frac{d}{dr}\left(\frac{\partial L}{\partial Z'}\right)
-\frac{\partial L}{\partial Z}=0~. \eea The first equation of
motion, the one for the function $Z(r)$, reads \be P\left(r
\frac{d Z}{dr}\right)^2+  M(Z) \left(r \frac{d Z}{dr}\right)+
Q(Z)=0~. \label{Zeq}
 \ee
If we algebraically solve the above equation we obtain the first
order differential equations \be \label{ZeqSol} r \frac{d
Z}{dr}=H(Z)~, \ee where $H(Z)$ are the solutions of second order
algebraic equation (\ref{Zeq}). For $P\neq 0:$ \be \label{solh}
H(Z)=\frac{-M(Z) + \sigma \sqrt{M(Z)^2-4 P Q(Z)}}{2P }~, \ee where
$\sigma$ is a sign; For $P=0$ we have only one solution: \be
H(Z)=-\frac{Q(Z)}{M(Z)}~. \ee Now we can derive the equation of
motion for the function $N(r)$. If we set
\be
 \bar{N}(r)=\sqrt{\frac{N(r)^2}{f(r)}}~,
\ee we obtain from the second Euler-Lagrange equation in
(\ref{lagrange}): \be \frac{d \bar{N}(r)}{d r} + \bar{C}(r)
\bar{N}(r)=0~, ~~~ \bar{C}(r) =\left[\frac{1}{r^4 G_1} \frac{d
(r^4 G_1)}{d r}-\frac{G_2}{rG_1}\right]~, \label{Neq} \ee where
\bea && G_1 = 2 P \left(r\frac{d Z}{d r}\right) +   M(Z)~, \\
&&G_2= M'(Z)\left( r\frac{d Z}{d r}\right) +  Q'(Z)~. \eea
Changing variables from $r$ to $Z$, Eq. (\ref{Neq}) becomes: \be
\frac{d \tilde{N}(Z)}{d Z} + \tilde{C}(Z) \tilde{N}(Z)=0~, ~~~
\tilde{C}(Z) =\frac{1}{H(Z)}\left[4
-\frac{\tilde{G}_2}{\tilde{G}_1}\right]+\frac{1}{\tilde{G}_1}\frac{d
\tilde{G}_1}{dZ} \label{NeqZ}~, \ee where \bea \label{gZ1} &&
\tilde{G}_1(Z) = 2 P H(Z) + M(Z)~, \nonumber \\ &&\tilde{G}_2(Z)=
M'(Z)H(Z) + Q'(Z)~. \eea

Finally, we emphasize that the parameter $\lambda$ does not appear
in the equations of motion, so they only depend on three
parameters: $\eta$, $\varrho$ and the cosmological constant
$\eta_{0a}$. This is similar to the 4D case
\cite{Kiritsis:2009rx}. The reason is that we are looking for
static solutions, hence extrinsic curvature terms do not
contribute to the equations of motion, as they contain only time
derivatives of the metric components. In addition, note that the
parameter $\lambda$ appears only in the extrinsic curvature part
of the action. Therefore, the solutions we will obtain in the
following sections will be valid for arbitrary values of
$\lambda$.

\section{Spherically symmetric solutions, special cases}

In this section we study static spherically symmetric solutions
for three special cases of the free coupling parameters  $\eta$
and $\varrho$: a) $\eta=0$ and $\varrho = 0$, b) $\eta=0$ and
$\varrho \neq 0$, c)  $\varrho=0$ and $\eta \neq 0$.

\subsection{No quadratic terms, $\eta=0$ and $\varrho = 0$}

If we set $\eta=\varrho=0$ in Eq. (\ref{Zeq}) we find \be
r\frac{dZ}{dr}=\frac{\eta_{0a}}{3}-4 Z~, \ee from which it follows
that \be -3  Z+\frac{\eta_{0a}}{4}+\frac{\tilde{C}_{\mu}}{r^4}=0~,
\label{Zeq1} \ee or equivalently we take the simple solution \be
f(r)=k+r^2 Z=k+\frac{\eta_{0a}}{12}r^2+\frac{\tilde{C}_{\mu}}{3
r^2}~, \ee where $\tilde{C}_{\mu}$ is a constant of integration.
If we set \be
\Lambda_{eff}=-\eta_{0a}~,~~\mu=-\frac{\tilde{C}_{\mu}}{3}~, \ee
the above equation takes the well-known form \be
f(r)=k-\frac{\Lambda_{eff}}{12} r^2-\frac{\mu}{r^2}~, \ee which is
the standard $AdS_5$ (for $\Lambda_{eff}<0$) or $dS_5$ (for
$\Lambda_{eff}>0$) or asymptotically flat (for $\Lambda_{eff}=0$)
Schwarzschild  black hole solution of 5D general relativity with a
cosmological constant.

\subsection{ $\eta=0$ and $\varrho \neq 0$}

If $\eta=0$ Eq. (\ref{PMQ}) implies $P=0$, hence Eq. (\ref{Zeq})
can be written as \be r\frac{dZ}{dr}=-\frac{Q(Z)}{M(Z)}~.
\label{Zeqfirst} \ee In this case the function $Q(Z)$ and $M(Z)$
can be written as \be\label{MQ} M(Z)=-3+6\varrho Z~,\ee \be
Q(Z)=\eta_{0a}-12 Z+12\varrho Z^2~, \label{MQ1} \ee and Eq.
(\ref{Zeqfirst}) becomes:
\be
r\frac{dZ}{dr} = -\frac{\eta_{0a}-12 Z+12 \varrho Z^2}{-3+6
\varrho Z}~. \ee Integration of this equation yields: \be 3\varrho
Z^2-3 Z+\frac{\eta_{0a}}{4}+\frac{\tilde{C}_{\mu}}{r^4}=0
\label{Zeq0}~, \ee where $\tilde{C}_{\mu}$ is an integration
constant which is related to the mass of the black hole. The
algebraic equation (\ref{Zeq0}) gives two solutions \be
Z(r)=\frac{1}{2 \varrho}+ \sigma \frac{\sqrt{3 (3 - \varrho
\eta_{0a}) r^4-12 \varrho \tilde{C}_{\mu}}}{6 \varrho~r^2}~, \ee
where $\sigma$ is a sign ($\sigma=\pm 1$), so for the function
$f(r)=k+r^2 Z$ we obtain \be \label{as1} f(r)=k+\frac{r^2}{2
\varrho}\left[1+ \sigma \sqrt{\left(1 - \frac{\varrho
\eta_{0a}}{3}\right) -\frac{4 \varrho \tilde{C}_{\mu}}{3
r^4}}\right]~. \ee In what follows we will assume that $\varrho
\tilde{C}_{\mu}<0$, because for $\varrho \tilde{C}_{\mu}>0$ the
range of radius $r$ has a lower bound $(r>r_{min}).$ This case
will be discussed further in section 5.2.

From the above equation (\ref{as1}) we can extract the large
distance asymptotic behaviour for $f(r)$, which reads \be
f(r)=k+\frac{1}{2 \varrho}\left(1 +\frac{\sigma }{\sqrt{3}}\sqrt{3
-  \varrho n_{0a}}\right) r^2-\frac{\sigma
}{\sqrt{3}}\frac{\tilde{C}_{\mu}}{
 \sqrt{3 -  \varrho n_{0a}}~ r^2}+O\left(\frac{1}{r^6}\right)~.
\label{asypt1} \ee Note that $f(r)$ has a large distance limit
only if $3> \varrho n_{0a}$, or else there is an upper bound for
the radius $r$. The asymptotic formula (\ref{asypt1}) is of the
form
\be
 f(r)\simeq k-\frac{\Lambda_{eff}}{12} r^2-\frac{\mu }{r^2}~,
\ee with \be \Lambda_{eff}=-\frac{6}{ \varrho}\left[1
+\frac{\sigma}{\sqrt{3}}\sqrt{3 -   \varrho
n_{0a}}\right],~\mu=\frac{\sigma}{\sqrt{3}}
\frac{\tilde{C}_{\mu}}{
 \sqrt{3 - \varrho n_{0a}}}~.
\ee Depending on the values of the free parameters $\varrho$ and
$\eta_{0a}$, the asymptotic behavior is either $AdS_5$ (for
$\Lambda_{eff}<0$), or $dS_5$ (for $\Lambda_{eff}>0$), or flat
(for $\Lambda_{eff}=0$). Also, note that $[\mu]=2$. and
$\Lambda_{eff}$ is an effective 5D cosmological constant. The mass
parameter of the black hole, when $\eta=\varrho=0$, is $m=(8 \pi
G_5)^{-1}\Lambda_{eff}^{-2} \mu$.

The Euler-Lagrange equations for $N(r)$ yield \be \frac{d
\tilde{N}(Z)}{d Z} + \tilde{C}(Z) \tilde{N}(Z)=0~, ~~~
\tilde{C}(Z) =\frac{1}{H(Z)}\left[4
-\frac{\tilde{G}_2}{\tilde{G}_1}\right]+\frac{1}{\tilde{G}_1}\frac{d
\tilde{G}_1}{dZ} \label{NeqZ1}~, \ee where
$$H(Z)=-\frac{Q(Z)}{M(Z)}~, \ \tilde{G}_1 =  M(Z)~,\ \tilde{G}_2 =
-\frac{Q(Z) M'(Z) }{M(Z)} +  Q'(Z)~.$$ Then we obtain
$$\tilde{C}(Z)=\frac{Q'(Z)-4M(Z)}{Q(Z)}=0~,$$ where we have taken
into account equations (\ref{MQ}), (\ref{MQ1}) for $M(Z)$ and
$Q(Z)$. Finally, we find
\begin{equation}
N(r)^2=f(r)~.
\end{equation}

\subsection{Comparing with the 5D Gauss-Bonnet gravity}

It is worth noting that the spherically symmetric solutions we
obtained in the previous section, for $\eta=0$ and $\varrho \neq
0$, are identical with the corresponding solutions of the 5D
relativistic Gauss-Bonnet (GB) gravity. The action of the GB
gravity is given by: \be \label{actionGB}
 S=\frac{1}{16 \pi G_5}\int d^D x \sqrt{|g^{(D)}|}
 \left \{ R^{(D)}-2 \Lambda+\hat{a} \hat{G}\right\},
 ~~D=5~,
\ee where the GB density $\hat{G}$ is \be
\hat{G}=R^{(5)abcd}R^{(5)}_{abcd}-4R^{(5)ab}R^{(5)}_{ab}+\left(R^{(5)}\right)^2,~~~
a,b,c,d=0,1,...4 \ee Note that in the GB gravity, the definition
of the symbols $R^{(5)}$, $R^{(5)ab}$ and $R^{(5)abcd}$ is based
on the relativistic  5D metric $g^{(5)}_{ab}$, where
$a,b,c,d=0,1,...4$. In addition, $\hat{a}$ is the Gauss Bonnet
coupling and $\Lambda$ is the 5D cosmological constant. The static
spherically symmetric solutions of the GB gravity in $AdS$ space,
for $D=5$, are of the form \be \label{metricGB} ds^2=-f(r)
dt^2+f^{-1}(r)~dr^2 +r^2 d\Omega_{k}^2 \ee  and \be\label{GBblack}
f(r)=k+\frac{r^2}{4 \hat{a}}\left[1+\sigma
\sqrt{1-8\hat{a}n^2+8\frac{a \mu_g}{r^4} }\right], ~\sigma=\pm 1~,
\ee in which $\mu_g$ is a constant of integration which is related
with the mass of the black hole, and the parameter $n^2=-2
\Lambda$ corresponds to a negative bulk cosmological constant.

If we replace \be
\hat{a}\rightarrow\frac{\varrho}{2},~n^2\rightarrow\frac{\eta_{0a}}{12},
~\mu_g\rightarrow-\frac{C_{\mu}}{3}\ee in equation
(\ref{GBblack}), we recover the black hole solution of Eq.
(\ref{as1}), for the specific case $\eta=0$ and $\varrho \neq 0$
of the previous section.

Note, that the condition $\eta=3 \eta_{2a} +  \eta_{2b}+
\eta_{2c}=0$  is satisfied in the case of GB coefficient
$\eta_{2a}= \hat{a}$, $\eta_{2b}=-4\hat{a}$ and
$\eta_{2c}=\hat{a}$, but there are other different combinations of
the coupling parameters $\eta_{2a},~  \eta_{2b},~ \eta_{2c}$ which
give $\eta=0$ and $\varrho\neq 0$. This is a very interesting
result which merits further investigation. Note also that the
relation $1 - \frac{\varrho \eta_{0a}}{3}=0$ corresponds to the
Chern-Simons limit of GB gravity.

\subsection{ $\varrho=0$ and $\eta \neq 0$}

\begin{figure}[h]
\begin{center}
\includegraphics[width=0.8 \textwidth, angle=0]{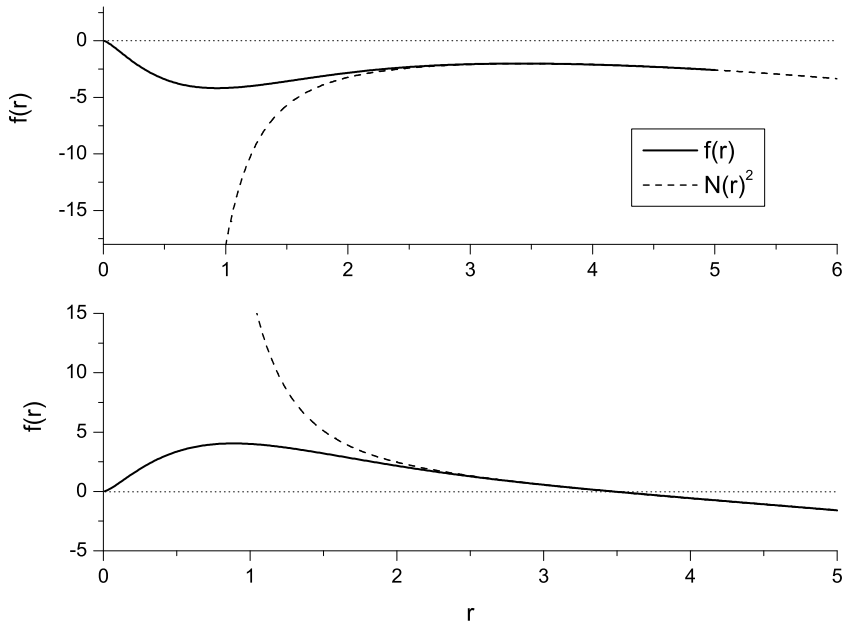}
\end{center}
\caption{\small A typical plot of $f(r)$ and $N(r)^2$ versus $r$,
in the case of $\varrho=0$ and $\eta \neq 0$, for $k=0$, $
\eta_{0a}=-1$ (positive cosmological constant),
$\tilde{C}_\mu=10$, $\eta=-0.1$ (top) and $\eta=+0.1$
(bottom)}\label{2}
\end{figure}

If we assume that $\varrho=0$, Eqs. (\ref{PMQ}) yield
\begin{eqnarray}\label{PMQ2}
&&P=3 \eta~,~~M(Z)=-3~, ~~Q(Z)=-12 Z+\eta_{0a}~.
\end{eqnarray}
By solving Eq. (\ref{Zeq}) we obtain the following two solutions
\be r\frac{dZ}{dr}=\frac{3-\sigma \sqrt{9-12\eta\eta_{0a}+144 \eta
Z}}{6\eta}~, \ee with $\sigma=\pm1$. The above differential
equation can be integrated analytically, to give \be \label{caseB}
\left(\frac{\sigma}{3}\sqrt{9-12\eta\eta_{0a}+144 \eta Z}-1\right)
e^{\left[\frac{\sigma}{3}\sqrt{9-12\eta\eta_{0a}+144 \eta
Z}-1\right]}=\frac{\tilde{C}_{\mu}}{r^4}~, \ee where
$\tilde{C}_{\mu}$ is an integration constant.

In this section we are interested only in solutions with
$\sigma=1$ and $\tilde{C}_{\mu}>0$. The other cases lack a short
distance limit (for details see Appendix A) and hence their
interpretation is problematic, as we explain in section 5.2 below.

From Eq. (\ref{caseB}) we obtain \be \label{caseBs}
\frac{1}{3}\sqrt{9-12\eta\eta_{0a}+144 \eta
Z}-1=W_{L}\left(\frac{\tilde{C}_{\mu}}{r^4}\right)~, \ee Note that
$W_{L}(x)$ is the Lambert function, which is defined as the real
solution of the equation $e^{W_{L}(x)} W_{L}(x)=x$. We recall some
of the properties of this equation: (a) for $x<-1/e$ it has no
real solutions, b) for $-1/e\leq x<0$ it has two real solutions
and, c) for $x\geq 0$ it has a unique real solution. In the case
b) we define the function $W_{L}(x)$ by demanding that $-1\leq
W_{L}(x)<0$ (the other set of solutions, which lies in the range
$(-\infty, -1),$ is not considered).

Now, from Eq. (\ref{caseBs}) we find  that \be \label{Z(r)2}
Z(r)=\frac{\eta_{0a}}{12}+\frac{1}{16\eta}\left(W_{L}^2
\left(\frac{\tilde{C}_{\mu}}{r^4}\right)+2
W_{L}\left(\frac{\tilde{C}_{\mu}}{r^4}\right)\right)~, \ee so we
find for the function $f(r)=k+r^2 Z(r):$ \be \label{f(r)2}
f(r)=k+\frac{\eta_{0a}}{12}r^2+\frac{r^2}{16\eta}\left(W_{L}^2
\left(\frac{\tilde{C}_{\mu}}{r^4}\right)+2
W_{L}\left(\frac{\tilde{C}_{\mu}}{r^4}\right)\right)~. \ee

The large $r$ asymptotic behaviour of Eq. (\ref{f(r)2}) is found
to be \be\label{asymf(r)2}
f(r)=k+\frac{\eta_{0a}}{12}r^2+\frac{\tilde{C}_{\mu}}{8 \eta
r^2}+O\left(\frac{1}{r^6}\right)~, \ee which is of the standard
form \be f(r)\simeq
k-\frac{\Lambda_{eff}}{12}r^2-\frac{\mu}{r^2}~, \ \
\Lambda_{eff}=-\eta_{0a}, ~~\mu=-\frac{\tilde{C}_{\mu}}{8 \eta}~.
\ee Now, Eqs. (\ref{PMQ}), (\ref{NeqZ}) and (\ref{gZ1}) yield \be
\tilde{C}(Z) =\frac{72 \eta }{9-12 \eta \eta_{0a}+144 \eta Z}
-\sigma \frac{24 \eta }{\sqrt{9-12 \eta \eta_{0a}+144 \eta Z}}
\label{NeqcaseB}~, \ee where the function $\tilde{N}(Z)$ satisfies
the equation \be \frac{d \tilde{N}(Z)}{d Z} + \tilde{C}(Z)
\tilde{N}(Z)=0~. \ee Solving this equation and choosing the
constant of integration appropriately, we get \be \label{caseB2}
\tilde{N}(Z)=\frac{3
e^{\left[\frac{\sigma}{3}\sqrt{9-12\eta\eta_{0a}+144 \eta
Z}-1\right]}}{\sqrt{9-12\eta\eta_{0a}+144 \eta Z}}~. \ee Note that
for $\sigma=1$ and $\tilde{C}_{\mu}>0$, if we take into account
Eqs. (\ref{caseBs}) and (\ref{caseB2}), the function $N(r)^2$ can
be expressed in the following closed form \be N(r)^2=f(r)
\tilde{N}(Z(r))^2=\frac{\tilde{C}_{\mu}^2 f(r) }{r^8
\left(W_{L}^2\left(\frac{\tilde{C}_{\mu}}{r^4}\right)+
W_{L}\left(\frac{\tilde{C}_{\mu}}{r^4}\right) \right)^2}~. \ee In
the large $r$ regime we find, from the above equation, that: \be
\label{asymN} N(r)^2=f(r)~\left(1+\frac{\tilde{C}_{\mu}^2 }{
r^8}+O\left(\frac{\tilde{C}_{\mu}^3}{r^{12}}\right)\right)~, \ee
hence in the large distance limit we recover the standard
asymptotic behavior $N(r)^2\simeq f(r).$

In Fig. \ref{2} we give a typical plot of the functions $f(r)$ and
$N(r)^2$  for zero spatial curvature $k=0$ and positive effective
cosmological constant $\Lambda_{eff}=1$. We see that $f(r)$ is
finite (in particular $f(0)=0$) but $N(r)^2$ blows up at the
singularity $r_s=0$. We also see that only when the coupling
$\eta$ is positive there exists a horizon, while for negative
$\eta$ there is a naked singularity. In the case of non-zero
spatial curvature the horizon $r_h$ is determined by solving
equation $f(r_h)=-k, ~(k=\pm1).$ It is possible to see by
inspection of Fig. \ref{2}, that for $k=\pm 1$, we have the same
situation which holds for zero spatial curvature. Finally, we
observe that $f(r)$ and $N(r)^2$ tend rapidly to their common
asymptotic behavior $(dS_5)$, as it is expected from the analysis
above (see Eqs. (\ref{asymf(r)2}) and (\ref{asymN})).

\section{Static solutions in the generic case ($\eta\neq 0$ and $\varrho\neq 0$)}

In the generic case, $\eta\neq 0$ and $\varrho\neq 0$, we obtain
from Eqs. (\ref{PMQ}), (\ref{ZeqSol}) and (\ref{solh})
\begin{equation}\label{genericEqZ}
r\frac{dZ}{dr}=\frac{3-6 \varrho Z+\sigma \sqrt{9-12 \eta_{0a}
\eta+ 36(\varrho-4\eta)(\varrho Z^2-Z)}}{6\eta}~.
\end{equation}
If we make the replacement
\be
Z=\frac{1}{2\varrho}-\frac{y}{3}
\ee
in Eq. (\ref{genericEqZ}), we find
\begin{equation} \label{genericEqy}
r\frac{dy}{dr}=\tilde{H}(y)~, ~~\tilde{H}(y)=-\frac{1}{\eta}
\left(\varrho y+\sigma\sqrt{A+ B y^2}\right)~,
\end{equation}
where the new parameters $A$ and $B$ are defined through \be A
\equiv -3 \eta_{0a} \eta+\frac{9\eta}{\varrho}~, ~~ B \equiv
\varrho (\varrho-4 \eta)~. \ee For the computation of the function
$N(r)$, Eqs. (\ref{PMQ}), (\ref{NeqZ}) and (\ref{gZ1}) yield \be
\frac{d \tilde{N}(y)}{d y} - \tilde{C}(y) \tilde{N}(y)=0~, ~~~
\label{NeqZgc} \ee where $\tilde{C}(y)$ is given by the equation
\be \label{genericEqNy}
 \tilde{C}(y)=-\frac{B}{\varrho} ~\frac{\varrho y+\sigma \sqrt{A+B y^2}}{A+B
 y^2}~.
 \ee
In the following sections we analyze two cases for the parameter
$B$: 1) $B\geq 0$, 2) $B<0$. The mathematical details for the
derivation of the final formulae for $f(r)$ and $N(r)$ are given
in Appendix B.

\subsection{$B\geq0$}

By integrating Eq. (\ref{genericEqy}) for $B\geq 0$ we obtain \be
\label{gcy1} \left|\varrho y+\sigma\sqrt{A+B y^2}\right|
\left|\sqrt{B} y+\sigma\sqrt{A+ B y^2}\right|^{-\frac{ \sqrt{B}}{
\varrho}}=\frac{\tilde{C}_{\mu}}{r^4}~. \ee Note, that in this
case  the parameter $A$ can be positive, negative or zero. For the
function $\tilde{N}$, we obtain from Eqs. (\ref{genericEqNy}) and
(\ref{NeqZgc}): \be \label{gcn1}
\tilde{N}(y)=\frac{\tilde{C}_{N}}{\sqrt{A+B y^2}\left|\sqrt{B}
y+\sigma\sqrt{A+ B y^2}\right|^{\frac{ \sqrt{B}}{ \varrho}}}~, \ee
where $\tilde{C}_{\mu}$ and $\tilde{C}_{N}$ are constants of
integration. Note, that $\tilde{C}_{\mu}$ must be always positive
($\tilde{C}_{\mu}>0$). Also, the constant $\tilde{C}_{N}$ is fixed
if we demand $\tilde{N}\rightarrow 1$ for $r\rightarrow +\infty$,
as we will discuss later in this section. An alternative
expression for $\tilde{N}$ can be found, if we take into account
Eqs. (\ref{gcy1}) and (\ref{gcn1}). In particular, we obtain that
\be \label{gcn2} \tilde{N}(y)=\frac{\bar{C}_{N}}{r^4\sqrt{A+B
y^2}\left|\varrho y+\sigma\sqrt{A+B
y^2}\right|}~,~~\bar{C}_{N}=\tilde{C}_{N} \tilde{C}_{\mu}~. \ee
For $B>0$, two cases for the ratio $\sqrt{B}/\varrho$ will be
discussed\footnote{Note that if $|\sqrt{B}/\varrho|=1$ we have no
solution at all.}: a) $|\sqrt{B}/\varrho|<1$ and b)
$|\sqrt{B}/\varrho|>1$.  The case $B=0$ will also be examined
separately in section 5.3.

\subsubsection{$|\sqrt{B}/\varrho|<1$ and $A> 0$ }

\begin{figure}[h]
\begin{center}
\includegraphics[width=0.8 \textwidth, angle=0]{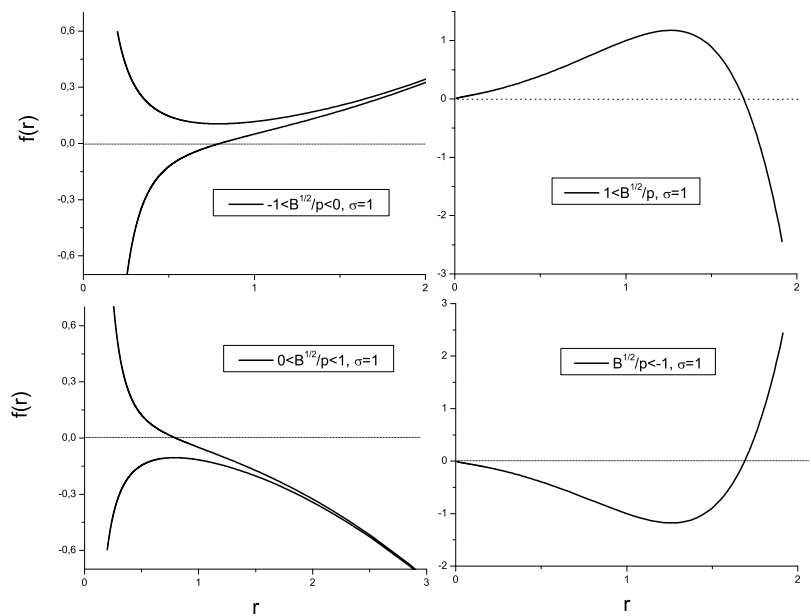}
\end{center}
\caption{\small A typical plot of $f(r)$ versus $r$, in the case
of $k=0, A=1, B=1, \tilde{C}_\mu=1, \sigma=1.$ In the left part of
the figure the parameter $\varrho$ equals $-10.0$ (top), or
$+10.0$ (bottom), while for the right part of the figure
$\varrho=+0.5$ (top) and $\varrho=-0.5$ (bottom)} \label{3}
\end{figure}

For $|\sqrt{B}/\varrho|<1$ and $A>0$, we can verify that Eq.
(\ref{gcy1}) above,  has two solutions for $y$ ($y_1$ and $y_2$)
for a given value of the radius $r$ in the range $[0,+\infty)$.
Hence, the function $f(r)$ has two branches $f_1(r)$ and $f_2(r)$,
which are exhibited in the left part of Fig. \ref{3}, when
$\sigma=1$ for two typical values of $\varrho$ ($\varrho=\pm 10$).
However, only one of them has a horizon for zero spatial curvature
($k=0$), while the other represents a spherically symmetric
solution with a naked singularity. In addition, as we see in Fig.
\ref{3}, for $\varrho>0$ the black hole solution is $AdS_5$
asymptotically, while for $\varrho<0$ is $dS_5$ (note that
$\varrho$ should be non-zero in the case we examine here).

In what follows, we will assume that $A>0$, as for negative $A$
the solutions have no large distance  limit \footnote{For $A<0$
and $|\sqrt{B}/\varrho|<1$ the left hand side of Eq. (\ref{gcy1})
is never zero, so the radius $r$ has an upper bound.} when
$|\sqrt{B}/\varrho|<1$, hence they will not be examined here.

In particular for $|\sqrt{B}/\varrho|<1$, if we take into account
Eq. (\ref{gcy1}), we obtain
\begin{itemize}
  \item $r \rightarrow +\infty \Rightarrow y\rightarrow y_0$ (large distance asymptotic behaviour)
  \item  $r \rightarrow 0 \Rightarrow y\rightarrow \pm \infty$ (short distance asymptotic behaviour)
\end{itemize}
where the plus and minus signs above correspond to the two
branches of the function f(r). In addition, \be \label{y01}
y_0=-sgn\left(\frac{\sigma}{\varrho}\right)
\sqrt{\frac{A}{\varrho^2-B}}~ \ee is the unique solution of the
following equation \be \varrho y_0+\sigma\sqrt{A+B y_0^2}=0~.
\label{y01a} \ee If we expand Eq. (\ref{gcy1}) around $y_0$ we can
find the large distance asymptotic behavior for $y(r)$ which reads
\be \label{asymy} y(r)\simeq y_0\mp\frac{3
\mu}{r^4}+O\left(\frac{1}{r^8}\right),~~\mu=|\varrho|\frac{|y_0
(\sqrt{B}-\varrho)|^{\frac{\sqrt{B}}{\varrho}}}{3(\varrho^2-B)}\tilde{C}_{\mu}~,
\ee hence \be f(r)=k+ r^2
Z=k+r^2\left(\frac{1}{2\varrho}-\frac{y}{3}\right) \simeq
\label{asympt} f(r)\simeq
k+\left(\frac{1}{2\varrho}-\frac{y_0}{3}\right) r^2 \pm
\frac{\mu}{r^2}+O\left(\frac{1}{r^{6}}\right)~, \ee which has the
standard asymptotic behaviour, $AdS_5$, $dS_5$ or flat, depending
on the values of the free parameters of the model. The plus and
minus signs in the above asymptotic formulas give rise to the two
branches of the function $f(r)$.

Now, for the function $\tilde{N}(y)$, if take into account Eqs.
(\ref{gcn2}) and (\ref{asymy}), we obtain the following large
distance asymptotic behaviour \be \label{asymy0}
\tilde{N}(y(r))\simeq1+O\left(\frac{1}{r^8}\right)~, \ee where the
constant of integration $\tilde{C}_{N}$ has been set to \be
\tilde{C}_{N}=- \sigma \varrho ~y_0 |y_0
(\sqrt{B}-\varrho)|^{\frac{\sqrt{B}}{\varrho}}~, \ee in order to
satisfy the condition $\tilde{N}(y_0)=1$.

\begin{figure}[h]
\begin{center}
\includegraphics[width=0.8 \textwidth, angle=0]{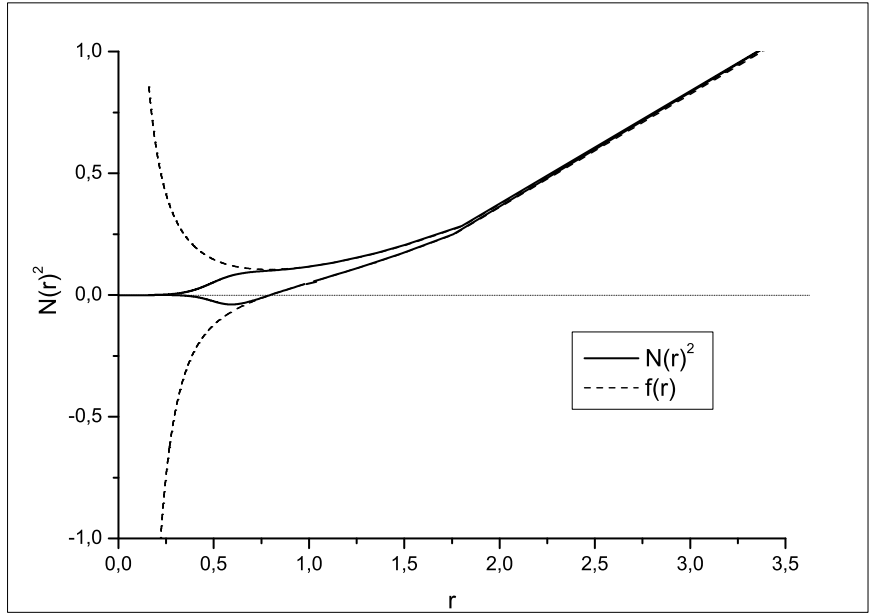}
\end{center}
\caption{\small A typical plot of the two branches of $N(r)^2$
versus $r$, in the case of $B>0$ and $|\sqrt{B}/\varrho|<1$, for
k=0, A=1, B=1, $\tilde{C}_\mu=1$, $\sigma=1$ and $\varrho$=10. We
observe that $N(r)^2\simeq f(r)$  when $r\rightarrow +\infty$, as
expected.}\label{4}
\end{figure}

The asymptotic behaviour of Eq. (\ref{asymy0}) is verified
graphically in Fig. \ref{4}. We observe that the function $N(r)^2$
tends rapidly to its asymptotic behaviour which is identical with
that of $f(r)$, as it is given by Eq. (\ref{asympt}). Also, we
would like to note, that the above analysis is valid only when $A>
0$. For negative $A$ we see that the left hand side of Eq.
(\ref{gcy1}) cannot vanish (see Eqs (\ref{y01}) and (\ref{y01a})
), so the radius $r$ has an upper bound. We conclude that this
class of solutions lacks physical interest, since the function
$f(r)$ has no large distance limit, so it will not be examined
here. The case $A=0$ is also examined separately in section 5.1.4.

\subsubsection{$|\sqrt{B}/\varrho|>1$ and $A>0$}

In the right part of Fig. \ref{3} we have plotted the function
$f(r)$ when $|\sqrt{B}/\varrho|>1$ ($\sigma=1$, $\varrho=\pm
0.5$). As we see, in contrast with the case
$|\sqrt{B}/\varrho|<1,$ now there is only one branch for the
function $f(r)$. However, there is an even more significant
difference, as this class of solutions does not exhibit the
standard $AdS_5$, $dS_5$ or flat asymptotic behaviour, as it is
shown in the following analysis. For negative $\sigma$
($\sigma=-1$) the behaviour of $f(r)$ does not change
significantly, so we do not give a figure for reasons of space.
The asymptotic behaviour of $f(r)$, for negative and positive
$\sigma$, is described in Eq. (\ref{asympt2}) below, and $f(r)$
obeys the same power law for both values of $\sigma$.

Although we will assume that $A>0$, there are also solutions for
negative $A$ in the case $|\sqrt{B}/\varrho|>1$ . These solutions
do not have any new features, as we comment in the following
section 5.1.3.

For $|\sqrt{B}/\varrho|>1$ and $A> 0$, from Eq. (\ref{gcy1}) above
we obtain
\begin{itemize}
  \item $r \rightarrow +\infty \Rightarrow y\rightarrow \epsilon (+\infty)$ (large distance asymptotic behaviour)
  \item  $r \rightarrow 0 \Rightarrow y\rightarrow -\epsilon (+\infty)$ (short distance asymptotic behaviour)
\end{itemize}
where $\epsilon$ is a sign defined as $\epsilon=sgn(\varrho
\sigma)$. Note, that the left hand side of Eq. (\ref{gcy1}) does
not vanish for a finite value of $y \ (y=y_{0}),$ the way it did
in the previous section.

For the large distance asymptotic behaviour of $y(r)$, if we take
into account Eq.  (\ref{gcy1}), we find: \be
\label{yasympt}y\simeq C_{\epsilon} ~
r^{\frac{4|\varrho|}{\sqrt{B}-|\varrho|}}~, \ee where the form of
the coefficient $C_{\epsilon}$ depends on the sign of the
parameter $\epsilon=sgn(\varrho \sigma)$, according to the
equations:
\begin{itemize} \label{asympt2}
  \item $C_{\epsilon}=\left(\frac{1}{2 \sqrt{B}}\right)^{\frac{\sqrt{B}}{\sqrt{B}-|\varrho|}}\left(\frac{\sqrt{B}+|\varrho|}{\tilde{C}_{\mu}}\right)^{\frac{|\varrho|}{\sqrt{B}-|\varrho|}}$, if $\epsilon=1$
  \item $C_{\epsilon}=-\left(\frac{2 \sqrt{B}}{ A}\right)^{\frac{\sqrt{B}}{\sqrt{B}-|\varrho|}}\left(\frac{\sqrt{B}+|\varrho|}{\tilde{C}_{\mu}}\right)^{\frac{|\varrho|}{\sqrt{B}-|\varrho|}}$, if $\epsilon=-1$
  \end{itemize}
The leading terms for $f(r)$ are \be \label{asym1} f(r)\simeq
k+\frac{r^2}{2\varrho}-\frac{c_{\epsilon}r^2}{3}~
r^{\frac{4|\varrho|}{\sqrt{B}-|\varrho|}}~, \ee which is not the
standard asymptotic behaviour for a 5D black hole solution, since
it is proportional to $r^{2+\delta}, \ \ \delta \equiv
~\frac{4|\varrho|}{\sqrt{B}-|\varrho|}>0~,$ rather than $r^2.$ The
large distance asymptotic behaviour for the modified lapse
function $\tilde{N}(r)$, if we take into account Eqs. (\ref{gcn2})
and (\ref{yasympt}), is \be \label{asym2} \tilde{N}(r)\simeq
\frac{\bar{C}_{N}}{C_{\epsilon}\sqrt{B}|\varrho+sgn(\varrho)
\sqrt{B}|}~~r^{-4 \frac{\sqrt{B}}{\sqrt{B}-|\varrho|}}~. \ee We
see that $\tilde{N}(r)\rightarrow 0$ for large $r$, which implies
a violation of the common large distance asymptotic behaviour
$N(r)^2 \simeq f(r)$. Finally, from the above equations
(\ref{asym1}) and (\ref{asym2}) we can determine the leading term
for the lapse function $N(r)$, which reads  \be N(r)\simeq
-\frac{\bar{C}_{N}}{3\sqrt{B}|\varrho+sgn(\varrho) \sqrt{B}|}~~r^{
\frac{-6 \sqrt{B}+2 |\varrho|}{\sqrt{B}-|\varrho|}}. \ee We
observe that also the lapse function $N(r)$ vanishes for large
values of $r$.

\subsubsection{$|\sqrt{B}/\varrho|>1$ and $A< 0$}

In fact, the solutions for $|\sqrt{B}/\varrho|>1$ and $A< 0$, is a
mixture  of solutions that were presented is the previous two
sections. More specifically every solution consists of two
branches. One of them has exactly the same short and large
distance asymptotic behaviour with the solutions of section 5.1.1.
This is mainly due to the fact that the left hand side of Eq.
(\ref{gcy1}) vanishes for a finite value of $y.$ The other branch
is similar to the solutions of section 5.1.2 but it lacks a short
distance limit. We will not discuss these cases further since
their behaviour is similar to the solutions we have already
discussed.

\subsubsection{$B>0$ and $A=0$ }

If we set \footnote{We remind the reader, that we can set $A=0$
only when $B$ is positive or zero.} $A=0$ (or
$\eta_{0a}\varrho=3$) in Eq. (\ref{gcy1}), we find \be
y(r)=C_{0}\; r^{\frac{4\varrho}{\left(\sqrt{B}-\varrho\right)}},
~~C_{0}=\sigma \left(\frac{1}{2
\sqrt{B}}\right)^{\frac{\sqrt{B}}{\sqrt{B}-\varrho}}\left(\frac{\varrho+\sqrt{B}}{\tilde{C}_{\mu}}\right)^{\frac{\varrho}{\left(\sqrt{B}-\varrho\right)}}~,
\ee so we obtain for the functions $f(r)$ and $\hat{N}(r):$ \be
f(r)=k+\frac{r^2}{2\varrho}-\frac{C_0}{3}~r^{2\frac{\sqrt{B}+\varrho}{\sqrt{B}-\varrho}}~
\ee and \be \hat{N}(r)=\frac{\bar{C}_{N}}{C_{0}^2
\sqrt{B}|\varrho+\sqrt{B}|}r^{-4
\frac{\sqrt{B}+\varrho}{\sqrt{B}-\varrho}}~. \ee As we see the
above solutions exhibit an unconventional asymptotic behaviour
which is not of the type $AdS_5$, $dS_5$ or flat 5D Schwarzschild
form.

\subsection{$B<0$}

\begin{figure}[h]
\begin{center}
\includegraphics[width=0.8 \textwidth, angle=0]{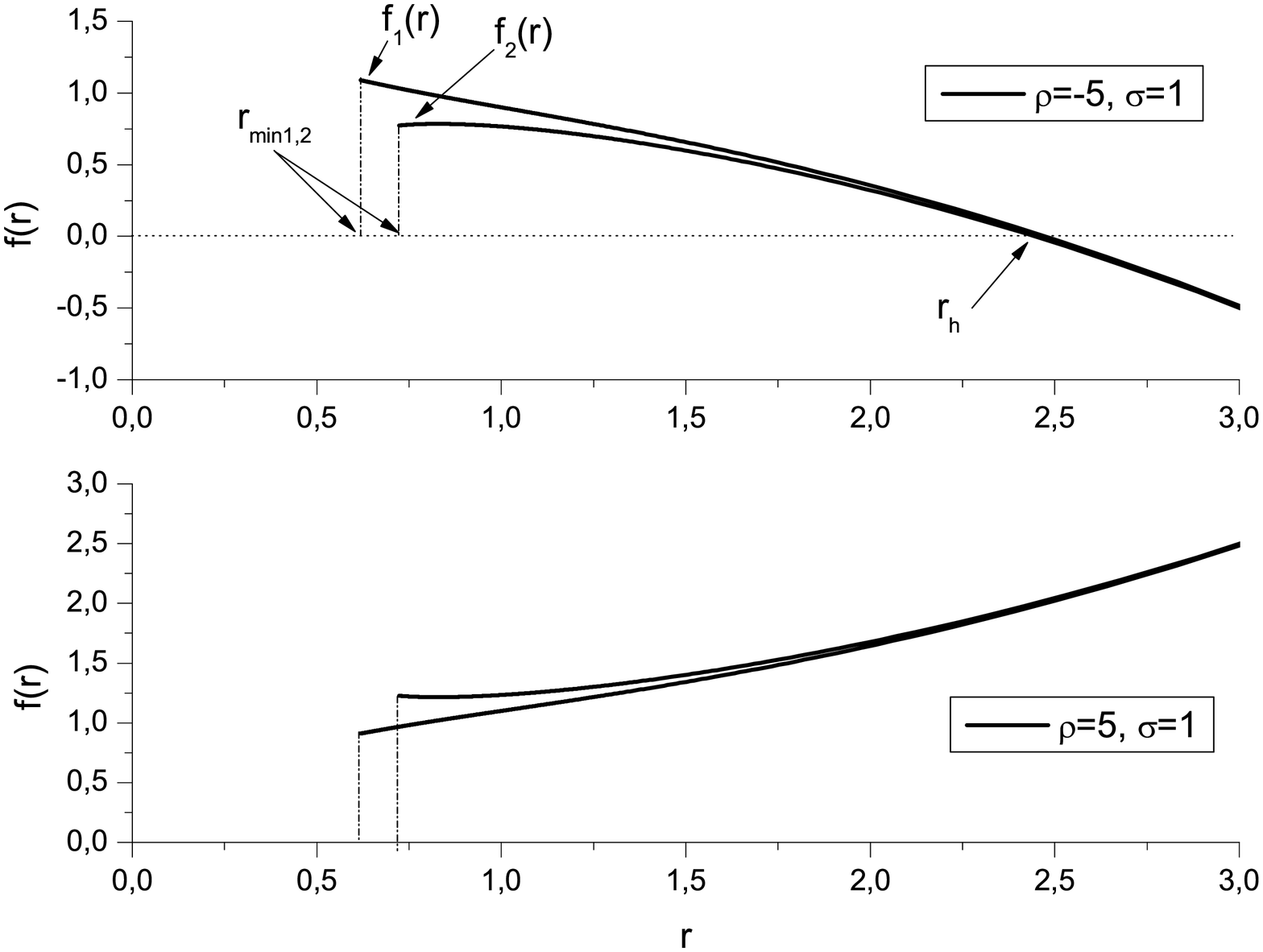}
\end{center}
\caption{\small $f(r)$ versus $r$, in the case of $B<0$, for k=0,
A=1, B=-1, $\tilde{C}_\mu=1$, $\sigma=1$ and $\varrho$=-5 (top),+5
(bottom) }\label{5}
\end{figure}

\begin{figure}[h]
\begin{center}
\includegraphics[width=0.8 \textwidth, angle=0]{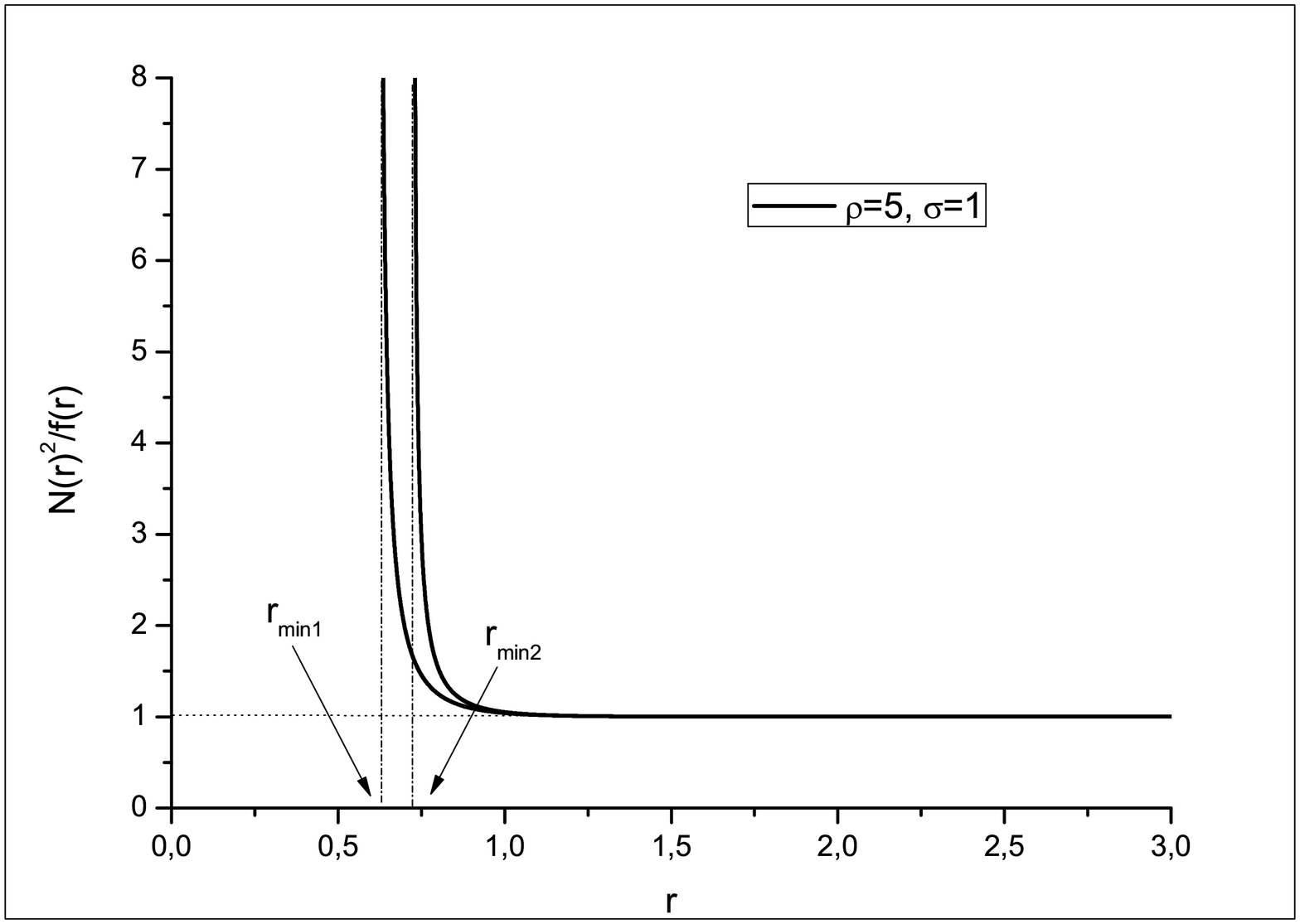}
\end{center}
\caption{\small $N(r)^2/f(r) $ versus $r$, in the case of $B<0$,
for k=0, A=1, B=-1, $\tilde{C}_\mu=1$, $\sigma=1$ and
$\varrho$=+5.} \label{6}
\end{figure}

\begin{figure}[h]
\begin{center}
\includegraphics[width=0.8 \textwidth, angle=0]{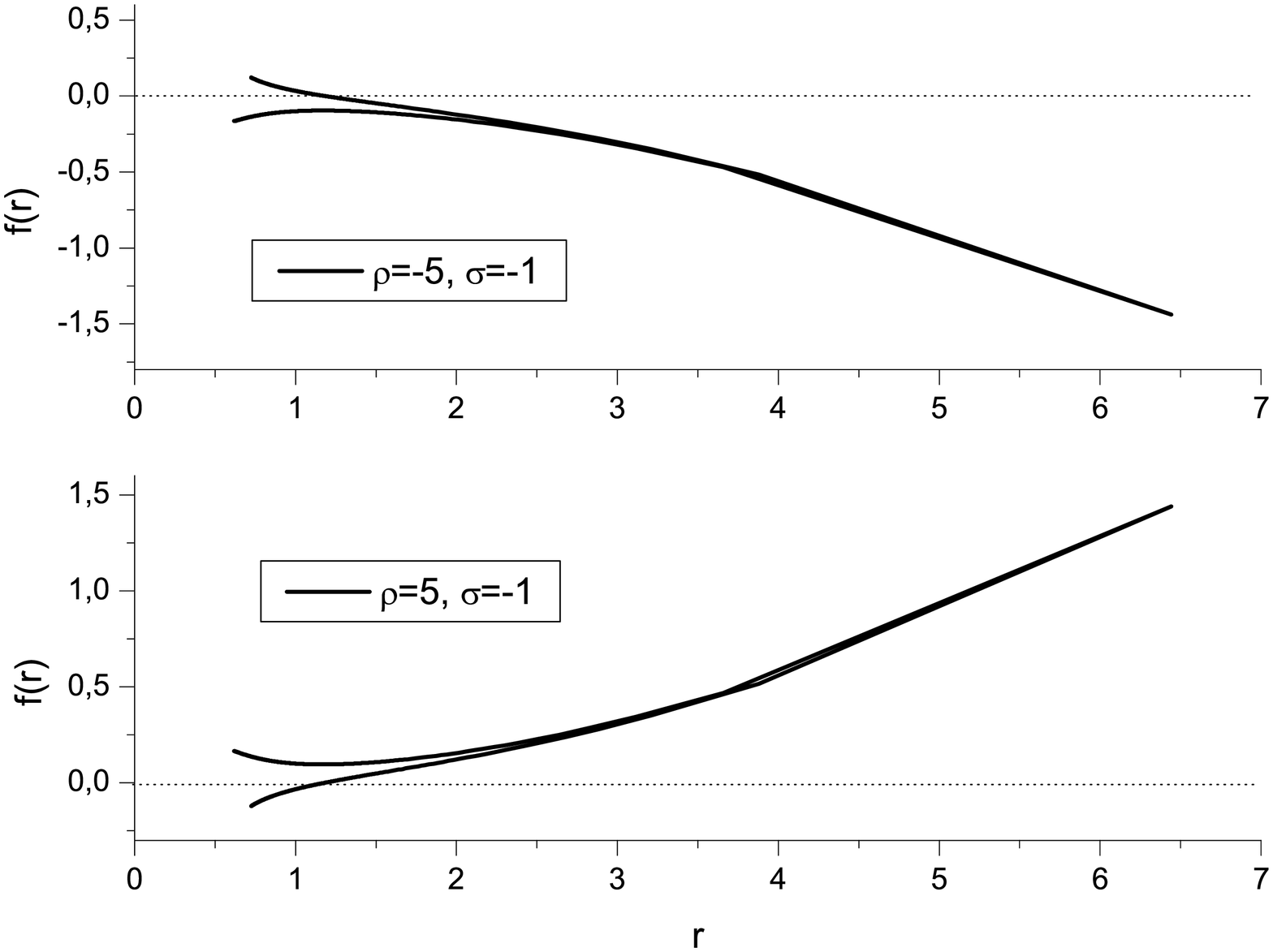}
\end{center}
\caption{\small $f(r)$ versus $r$, in the case of $B<0$, for k=0,
A=1, B=-1, $\tilde{C}_\mu=1$, $\sigma=-1$ and $\varrho$=-5.5
}\label{7}
\end{figure}

For $B<0$, if we integrate Eq. (\ref{genericEqy}),  we find \be
\label{gcy2} \left|\varrho y+\sigma\sqrt{A-|B| y^2}\right|e^{
\frac{\sigma  \sqrt{|B|}}{ \varrho}\tan^{-1}
\left(\frac{\sqrt{|B|} y}{\sqrt{A-| B| y^2}}\right)}
=\frac{\tilde{C}_{\mu}}{r^4}~. \ee Note that in this case the
parameter $A$ should be always positive ($A>0$). Eqs.
(\ref{NeqZgc}) and (\ref{genericEqNy}) yield: \be \label{gcn3}
\tilde{N}(y)=\frac{\tilde{C}_{N}}{\sqrt{A-|B| y^2}}e^{
\frac{\sigma  \sqrt{|B|}}{ \varrho}\tan^{-1}
\left(\frac{\sqrt{|B|} y}{\sqrt{A-| B| y^2}}\right)}~. \ee Now, if
we use Eqs. (\ref{gcy2}) and (\ref{gcn3}) we can get an
alternative expression for $\tilde{N}$ according to the equation
\be \tilde{N}(y)=\frac{\bar{C}_{N}}{r^4\sqrt{A-|B|
y^2}\left|\varrho y+\sigma\sqrt{A-|B| y^2}\right|}~. \ee If we
take into account the restriction from the square root in Eq.
(\ref{gcy2}), we find that $y$ varies in the finite interval
$|y|<\sqrt{\frac{A}{|B|}}$.

In Fig. \ref{5} we have plotted the function $f(r)$, for positive
$\sigma$ ($\sigma=1$),  two typical values of $\varrho$
($\varrho=5>0$ and $\varrho=-5<0$) and zero spatial curvature
$k=0$. As we see $f(r)$ consists of two distinct branches $f_1(r)$
and $f_2(r)$, with a $dS_5$ (for $\varrho<0$) and $AdS_5$ (for
$\varrho>0$) large distance asymptotic behaviours correspondingly.
In contrast with the previous case ($B>0$) the range of radius r
terminates at a lower non-zero bound $r_{min1}$ (for $f_1(r)$) and
$r_{min2}$ (for $f_2(r)$). For zero spatial curvature, $k=0,$ we
observe that for $\varrho<0$ both the two branches have a horizon,
while for $\varrho>0$ we do not obtain horizons at all. Notice
that the lower branch in the bottom of Fig. \ref{5} may have a
horizon if $k=-1.$ In Fig. \ref{6} we see that the function
$N(r)^2/f(r)$ tends to unity as $r$ tends to infinity, but
$N(r)^2/f(r)$ blows up when $r\rightarrow r_{min1,2}$. We observe
that in this case there is an infinite set of singularity points
which lie on a 4D hypersurface of constant radius $r_{min1,2}$.
(This situation is to be compared with the standard case of a
black hole solution where the singularity point is located at the
origin). These solutions may have a physical relevance in the case
where the singular shell is protected by a horizon, for example
the branches of $f(r)$ with negative $\varrho$ develop a horizon
as we pointed out above.

Also, in Fig. \ref{7}, we have plotted the function $f(r)$ for
negative $\sigma$  ($\sigma=-1$) for the same two typical values
of $\varrho$ and $k=0$ . Note that there are no significant
differences if we compare with the corresponding figure for
$\sigma=1$ (Fig. \ref{5}). However, we see that for negative
$\sigma$ only one of the two branches of $f(r)$ possesses a
horizon, while the other exhibits a naked singularity (in fact we
have a naked singular shell).

Although in the above mentioned figures we have used specific
values for the free parameters,  our conclusions are quite general
as the values we chose represent a wide range of the parameter
space.

In what follows we try to better understand some of the properties
of the solutions by using the analytical formulae of Eqs.
(\ref{gcy2}) and (\ref{gcn3}) above. First, we would like to
stress that the transcendental equation (\ref{gcy2}) gives two
solutions for $y$ for a fixed value of the radius $r$, as it may
also be seen in Figs. \ref{5} and \ref{6}. In particular the two
branches $f_1(r)$ and $f_2(r)$ correspond to the two intervals of
the parameter $y$: 1) $[-\frac{A}{|B|},y_0]$ and  2)
$[y_0,\frac{A}{|B|}]$. Now, if we take into account Eq.
(\ref{gcy2}) we find that for large $r$ the function $y(r)$ tends
to a constant value, or equivalently that
\begin{equation}
 r \rightarrow +\infty \Rightarrow y\rightarrow y_0~.
 \end{equation}
We note that \be y_0=-sgn\left(\frac{\sigma}{\varrho}\right)
\sqrt{\frac{A}{\varrho^2+|B|}} \ee is the unique solution of the
equation \be \varrho y_0+\sigma\sqrt{A-|B| y_0^2}=0~. \ee Note,
that for negative $B$ ($B<0$) there is no short distance limit
with  $r\rightarrow 0$. We find that $r>r_{min}$, where $r_{min}$
is a non-zero lower bound for the radius $r$, which is given by
the equation \be r_{min}= \left(\tilde{C}_{\mu}^{-1} |\varrho|
\sqrt{\frac{A}{|B|}}\right)^{-1/4} e^{\mp \frac{\sigma
\sqrt{|B|}\pi}{ 8\varrho}}~. \ee It turns out that:
\begin{equation}
\lim_{r \rightarrow r_{min}} y = \pm \sqrt{\frac{A}{|B|}}~.
\end{equation}
The large distance asymptotic behaviour for $y(r)$ is obtained by
expanding Eq. (\ref{gcy2}) around $y_0:$ \be \label{asymy2}
y(r)\simeq y_0\mp\frac{3
\mu}{r^4}+O\left(\frac{1}{r^8}\right),~~\mu=\frac{|\varrho|\tilde{C}_{\mu}}{3(\varrho^2+|B|)}
e^{\frac{\sqrt{|B|}}{\varrho}\tan^{-1}\left(\frac{\sqrt{|B|}}{{\varrho}}\right)}~,
\ee hence for the function $f(r)$ we find the corresponding
asymptotic behaviour \be f(r)\simeq
k+\left(\frac{1}{2\varrho}-\frac{y_0}{3}\right) r^2 \pm
\frac{\mu}{r^2}+O\left(\frac{1}{r^{6}}\right)~, \ee which has the
standard $AdS_5$, $dS_5$ or flat asymptotic behaviour, depending
on the values of the free parameters of the model.

For the function $\tilde{N}(y(r))$, when the radius tends to
infinity, we find \be
\tilde{N}(y(r))\simeq1+O\left(\frac{1}{r^8}\right)~, \ee where the
constant of integration $\tilde{C}_{N}$ has been set to \be
\tilde{C}_{N}=- \sigma \varrho y_0
e^{-\frac{\sqrt{|B|}}{\varrho}\tan^{-1}\left(\frac{|B|}{\varrho}\right)}~,
\ee to satisfy the condition $\tilde{N}(y_0)=1$.

\subsection{$B=0$}

Although the special case of $B=0$ is included in section 5.1.1,
we present the relevant results separately here, since $f(r)$ can
be expressed explicitly as a function of $r$. If $B=0$ and $A>0$,
from Eq. (\ref{genericEqZ}) we obtain
\begin{equation} \label{genericB0}
r\frac{dy}{dr}=-\frac{1}{\eta} \left(\varrho
y+\sigma\sqrt{A}\right)~.
\end{equation}
From $B=\varrho (\varrho-4 \eta)=0$ and $\varrho \neq 0$, we
conclude that $\varrho=4 \eta$, hence the above equation
(\ref{genericB0}) can be written as
\begin{equation} \label{genericB01}
r\frac{dy}{dr}=-\frac{4}{\varrho} \left(\varrho
y+\sigma\sqrt{A}\right)~,
\end{equation}
which yields \be y=-\frac{\sigma \sqrt{A}}{\varrho}\mp
\frac{\tilde{C}_{\mu}}{\varrho r^4}~, \ee while for the function
$f(r)$ we find \be f(r)=k+\left(\frac{1}{2\varrho}+\frac{\sigma
\sqrt{A}}{3 \varrho}\right)r^2\pm \frac{\tilde{C}_{\mu}}{3 \varrho
r^2}~.\label{llbh} \ee This has the standard form of a $AdS_5$,
$dS_5$ or flat solution. The above formulae can also be obtained
if we set $B=0$ in Eqs. (\ref{y01}), (\ref{asymy}) and
(\ref{asympt}) of section 5.1.1. Finally, from Eqs. (\ref{NeqZgc})
and (\ref{genericEqNy}), choosing suitably the integration
constant, we find that \be N(r)^2=\frac{1}{f(r)}~. \ee

It is interesting to note that (at least the positive branch of)
the solution (\ref{llbh}) coincides, after a proper identification
of the parameters, with the black hole solutions found in
Lanczos-Lovelock theories \cite{Crisostomo:2000bb}. The
Lanczos-Lovelock action is a polynomial of degree [D/2] (the
integer part of D/2)  in the curvature and in 5D it has solutions
of the Schwarzschild-AdS type.

\section{Conclusions and discussion}

We studied static spherically symmetric solutions in the framework
of the 5D Horava-Lifshitz gravity. We considered an action
consisting of terms of up to second order in the curvature and we
solve the theory with a non-projectable spherically symmetric
ansatz for the metric. The black hole spectrum we found is
controlled by three parameters $\eta$, $\varrho$ and $\eta_{0a}$,
where $\eta_{0a}$ is a cosmological constant. The black hole
solutions we found do not depend on the parameter $\lambda$ which
measures the departure from Lorentz invariance as it  appears only
in the extrinsic curvature part of the action.

We presented a full analysis of 5D Horava-Lifshitz static
solutions scanning the values of the free parameters $\eta$,
$\varrho$ and $\eta_{0a},$ which can be positive, negative or
zero. This analysis comes as a generalization and extension of the
4D case studied in \cite{Kiritsis:2009rx}.  More specifically, we
obtained three main sets of solutions: the two special cases
$(\eta= 0, \ \varrho \neq 0), \ {\rm and} \ (\eta\neq 0, \ \
\varrho = 0)$ and the generic case $(\eta\neq 0,\ \ \varrho \neq
0).$ In all cases we obtained analytic black hole solutions which
have the standard $AdS_5$, $dS_5$ of flat asymptotic behaviour,
plus the well-known $1/r^2$ tail. However, we also obtained
solutions with an unconventional short and large distance
asymptotic behaviour. For example, in the generic case, we found
solutions with an asymptotic fall-off which is stronger than the
$AdS_5$ or $dS_5$ asymptotic behaviour. We have also found
solutions (in the cases with $\eta \neq 0)$ in which the
asymptotic behaviour is the usual one, but the radius has a lower
bound $r_{min}$ different from zero.  Also, in many cases we
obtained solutions with a naked singularity.

We also found static solutions which, after a proper
identification of coupling parameters, coincide with static black
hole solutions of relativistic gravity theories with quadratic
curvature correction terms. One class of these solutions consists
of the Schwarzschild-AdS  black hole solutions of five-dimensional
Lanczos-Lovelock gravity theories. Another class of solutions
contains the well-known Gauss-Bonnet black hole solutions. The
interesting result we obtained in our investigation is that the
non-relativistic solutions of the HL gravity corresponding to the
Gauss-Bonnet solutions can be obtained for various combinations of
the coupling parameters $\eta$ and $\varrho $ and not just the
standard Gauss-Bonnet combination. This may be attributed to the
fact that the HL static solutions are insensitive to the coupling
parameter $\lambda$, so they hold even if $\lambda \neq 1$ (the
value which signals the breaking of Lorentz invariance).

We do not have a full understanding of the quantum UV structure of
the Horava-Lifshitz gravity. The $\beta$ functions for the UV
marginal couplings have not yet been calculated, so the claim that
in the IR the $\lambda=1$ is a fixed point is an assumption. The
fact that a class of our solutions coincides with relativistic
Gauss-Bonnet black hole solutions is suggesting that in the
running of couplings towards the IR, the Gauss-Bonnet regime is
reached before the 5D gravity is attained. After all the
Gauss-Bonnet theory is an UV correction of 5D gravity.

Important issues remain to be investigated. One of them is the
stability of our solutions. For example, the stability of the
class of solutions we found which coincides with the Gauss-Bonnet
solutions is an interesting issue to be studied. The stability of
the Gauss-Bonnet static solutions has been extensively
studied~\cite{stability}. It was found in~\cite{Charmousis:2008ce}
that one branch of these solutions suffers from ghost-like
instability up to the strongly coupled Chern-Simons limit where
linear perturbation theory breaks down. In our case because the
Lorentz invariance is broken in the UV this behaviour could be
different. Therefore, a careful stability analysis is required.

Other interesting issues are the thermodynamic properties of our
solutions, or the contribution of terms of higher order in the
curvature, which have been omitted in the present work. It would
also be interesting to generalize the solutions we presented here
in the presence of electric charge.

Finally, one field that our findings can be applied  is the
extra-dimensional gravity theories, in particular the  brane world
models. Note that in contrast to the standard vacuum of
Randall-Sundrum \cite{Randall:1999ee}, the five-dimensional AdS
black hole vacuum does not preserve 4D Lorentz invariance on the
brane, which may have interesting phenomenological implications
\cite{Csaki:2000dm,Farakos:2008rv}. For appropriate ranges of the
coupling parameters, we have obtained solutions with an $AdS_5$
large distance asymptotic behaviour, plus the standard $1/r^2$
tail of a usual 5D Schwarzschild black hole. These solutions may
serve as backgrounds in the framework of brane worlds models, but
now there is additional advantage: the starting point is a
renormalizable theory such as the 5D HL gravity, in contrast with
the 5D General Relativity, which is non-renormalizable and
requires a UV complement.

\section{Acknowledgements}

We thank H. Kiritsis and G. Kofinas for stimulating discussions.

\section{Appendix A: Asymptotic behaviour analysis in the case $\varrho=0$ and $\eta\neq 0$}

In this appendix we will examine the short and large distance
asymptotic behaviour of the solution (\ref{caseB}) in the case
$\varrho=0$ and $\eta\neq 0$. In particular we will examine two
cases I) $\sigma=1$, and II)  $\sigma=-1$ for the equation \be
\label{caseB3} \left(\frac{\sigma}{3}\sqrt{9-12\eta\eta_{0a}+144
\eta Z}-1\right)
e^{\left[\frac{\sigma}{3}\sqrt{9-12\eta\eta_{0a}+144 \eta
Z}-1\right]}=\frac{\tilde{C}_{\mu}}{r^4}~. \ee As we will see, the
only interesting case is that for $\sigma=1$ and
$\tilde{C}_{\mu}>0$. The problem in other cases is that $Z(r)$ is
not defined on the whole interval $[0,+\infty)$.

\subsection{Case I ($\sigma=1$)}

Although, the case $\sigma=1$ and $\tilde{C}_{\mu}>0$ has been
examined in detail in section 4.4, we summarize  our results here
\begin{itemize}
  \item $r\rightarrow +\infty$ $\Rightarrow$ $Z \rightarrow \frac{\eta_{0a}}{12}$~~(and $Z>\frac{\eta_{0a}}{12}$)
  \item $r\rightarrow 0$ $\Rightarrow$ $Z\rightarrow sgn(\eta)(+\infty) $
\end{itemize}
For $\tilde{C}_{\mu}>0$, we see that $Z(r)$ is well defined in the
range $[0,+\infty)$.

For $\sigma=1$ and $\tilde{C}_{\mu}<0$ we find
\begin{itemize}
  \item $r\rightarrow +\infty$ $\Rightarrow$ $Z\rightarrow \frac{\eta_{0a}}{12}$~~(and $Z<\frac{\eta_{0a}}{12}$)
  \item $r\rightarrow 0$ $\Rightarrow$ the limit does not exist, as there is a lower bound $r>r_{min}$
\end{itemize}
For $\tilde{C}_{\mu}<0$, the function $Z(r)$ is defined in a range
of the form $[r_{min},+\infty)$, where $r_{min} > 0.$ In both
cases the large distance asymptotic behaviour is given by Eq.
(\ref{asymf(r)2}) in section 4.4, which is of the standard form
\be \label{sf} f(r)\simeq
k-\frac{\Lambda_{eff}}{12}r^2-\frac{\mu}{r^2}~. \ee In addition,
for $\tilde{C}_{\mu}<0$, the $r_{min}$ can be determined if we
take into account that the Lambert function has a lower bound
$W_{L}(x)\geq -1/e$, we then obtain \be \label{rmin}
r_{min}=\sqrt[4]{-e \tilde{C}_{\mu}}~, \ee and \be \label{Zmin}
Z(r_{min})=\frac{12\eta\eta_{0a}-9}{144 \eta}~. \ee In order to
derive the above equation we used that $W_{L}(-1)=-1/e$.

\subsection{Case II ($\sigma=-1$)}

For $\sigma=-1$, from Eq. (\ref{caseB3}) we see that the parameter
$\tilde{C}_{\mu}$ should be negative, hence we examine only the
case $\tilde{C}_{\mu}<0$
\begin{itemize}
  \item $r\rightarrow +\infty$ $\Rightarrow$ $Z \rightarrow sgn(\eta)(+\infty)$
  \item $r\rightarrow 0$ $\Rightarrow$ the limit does not exist, as there is a lower bound $r>r_{min}$
\end{itemize}
We see that for $\sigma=-1$, the function $Z(r)$ is defined in a
range of the form $[r_{min},+\infty)$, where $r_{min}$ and
$Z(r_{min})$ are given by Eqs. (\ref{rmin}) and (\ref{Zmin})
above. In addition, the large distance asymptotic behaviour in
this case can be estimated from Eq. (\ref{caseB3}) if we keep only
the exponential term \be Z(r)\simeq \frac{1}{16 \eta}
\ln\left(\frac{-\tilde{C}_{\mu}}{r^4}\right)~, \ee hence \be
f(r)\simeq k+ r^2\frac{1}{16 \eta} \ln\left(\frac{-e
\tilde{C}_{\mu}}{r^4}\right)~, \ee while for $\tilde{N}$, from Eq.
(\ref{caseB2}), we obtain \be
\tilde{N}(Z(r))\simeq\frac{\tilde{C}_{\mu}}{3 r^4
\ln\left(\frac{-e \tilde{C}_{\mu} }{r^4}\right)}~. \ee We conclude
that for $\sigma=-1$ (and $\varrho=0$, $\eta\neq 0$ ) the large
distance asymptotic behaviour is not of the standard form
(\ref{sf}), so we will not study this class of solutions further.

\section{Appendix B: Technical details in the generic case  ($\eta\neq 0$ and $\varrho\neq 0$)}

In this section we present same mathematical details for the
derivation of formulae of Eqs. (\ref{gcy1}) and (\ref{gcy2}). In
particular we have to integrate Eq. (\ref{genericEqZ})
\begin{equation}
r\frac{dZ}{dr}=\frac{3-6 \varrho Z+\sigma \sqrt{9-12 \eta_{0a}
\eta+ 36(\varrho-4\eta)(\varrho Z^2-Z)}}{6\eta}~.
\end{equation}
In order to perform the above integral we make the replacement
$Z=\frac{1}{2\varrho}-\frac{y}{3}$, then we obtain
\begin{equation}
r\frac{dy}{dr}=\tilde{H}(y)~, ~~\tilde{H}(y)=-\frac{1}{\eta}
\left(\varrho y+\sigma\sqrt{A+ B y^2}\right)~,
\end{equation}
where \be A=-3 \eta_{0a} \eta+\frac{9\eta}{\varrho}~, ~~ B=\varrho
(\varrho-4 \eta)~. \ee Note that $$\frac{\eta
\varrho}{2(\varrho^2-B)}=\frac{1}{8}~,$$ so we have $
\varrho^2-B=4\varrho\eta \neq 0$. For $B\neq 0$, we obtain \bea
\label{integ1} c+\ln(r)&=&-\eta\int\frac{ dy}{\varrho
y+\sigma\sqrt{A+ B y^2}}=-\eta\int dy ~~\frac{\varrho
y-\sigma\sqrt{A+ B y^2}}{(\varrho^2-B)y^2-A}\\ &=&
-\frac{1}{8}\ln\left|(\varrho^2-B)y^2-A\right|+\sigma \eta \int dy
~~\frac{\sqrt{A+ B y^2}}{(\varrho^2-B)y^2-A}~, \eea where $c$ is
an integration constant. In order to calculate the second integral
in Eq. (\ref{integ1}) separately, we write the corresponding
integrand in the form \bea \frac{\sqrt{A+ B
y^2}}{(\varrho^2-B)y^2-A}=\frac{1}{\varrho^2-B}\left(\frac{B}{\sqrt{A+
B y^2}}+\frac{A\varrho^2}{\sqrt{A+ B
y^2}}\frac{1}{\left((\varrho^2-B)y^2-A\right)}\right)~. \eea For
$B>0$ we obtain \bea \label{intB1}
 \int \frac{\sqrt{A+ B y^2}}{(\varrho^2-B)y^2-A}&= &\frac{\sqrt{B}}{\varrho^2-B} \ln\left|\sqrt{B} y+\sqrt{A+ B y^2}\right| \nonumber \\
&-&\frac{\varrho }{2(\varrho^2-B)}\ln\left|\frac{\varrho
y-\sqrt{A+ B y^2}}{\varrho y+\sqrt{A+ B y^2}}\right|~, \eea while
for $B<0$ we get \bea\label{intB2} \int dy \frac{\sqrt{A- |B|
y^2}}{(\varrho^2+|B|)y^2-A}&=&\frac{1}{\varrho^2+|B|}\sqrt{|B|}\tan^{-1}\left(\frac{\sqrt{|B|}
y}{\sqrt{A-|B| y^2}}\right) \nonumber \\ &-&\frac{\varrho
}{2(\varrho^2+|B|)}\ln\left|\frac{\varrho y-\sqrt{A+ B
y^2}}{\varrho y+\sqrt{A+ B y^2}}\right|~, \eea where we have taken
into account that \be \frac{A\varrho^2}{\sqrt{A+ B
y^2}}\frac{1}{\left((\varrho^2-B)y^2-A\right)}=-\frac{A\varrho^2}{(A+
B y^2)^{3/2}}\frac{1}{\left(1-\frac{\varrho^2
y^2}{A+By^2}\right)}~, \ee and the well known relations \be
\frac{d}{dy}\left(\frac{y}{\sqrt{A+B y^2}}\right)=\frac{A}{(A+ B
y^2)^{3/2}}~, ~~\int
\frac{dx}{1-x^2}=\frac{1}{2}\ln\left|\frac{1-x}{1+x}\right|~,~~
x=\frac{\varrho y}{\sqrt{A+B y^2}}~. \ee Moreover, we will use the
identity \be \label{ln1} \frac{1}{2}\ln\left|\frac{\varrho
y-\sqrt{A+ B y^2}}{\varrho y+\sqrt{A+ B y^2}}\right|
=-\frac{1}{2}\ln\left|(\varrho^2-B)y^2-A\right|+\ln\left|\varrho
y+\sqrt{A+B y^2}\right|~. \ee For $B>0$ from Eqs. (\ref{integ1}),
(\ref{intB1}) and (\ref{ln1}), if we take into account that $
\varrho^2-B=4\varrho\eta $, we obtain \bea c+\ln(r)&=&
\frac{\sigma-1}{8}\ln\left|(\varrho^2-B)y^2-A\right|+\frac{\sigma
\sqrt{B}}{4 \varrho} \ln\left|\sqrt{B} y+\sqrt{A+ B
y^2}\right|\nonumber\\ &-&\frac{\sigma}{4}\ln\left|\varrho
y+\sqrt{A+B y^2}\right|~, \eea or equivalently \be
c'+\ln(r)=-\frac{1}{4}\ln\left|\varrho y+\sigma\sqrt{A+B
y^2}\right|+\frac{\sqrt{B}}{4 \varrho} \ln\left|\sqrt{B}
y+\sigma\sqrt{A+ B y^2}\right|~,\nonumber \ee where $c'$ is a new
constant which is defined appropriately. Finally, we obtain \be
\left|\varrho y+\sigma\sqrt{A+B y^2}\right| \left|\sqrt{B}
y+\sigma\sqrt{A+ B y^2}\right|^{-\frac{ \sqrt{B}}{
\varrho}}=\frac{\tilde{C}_{\mu}}{r^4}~, \ee where
$\tilde{C}_{\mu}=e^{c'}$ is the final integration constant which
is related to the mass of the black hole. For $B<0$, in a similar
way, if we use Eq. (\ref{intB2}) instead of (\ref{intB1}), we
obtain \be \left|\varrho y+\sigma\sqrt{A-|B| y^2}\right| e^{
\frac{\sigma  \sqrt{|B|}}{ \varrho}\tan^{-1} \left(\frac{\varrho
y}{\sqrt{A-| B| y^2}}\right)} =\frac{\tilde{C}_{\mu}}{r^4}~. \ee
Finally, for the computation of $N(r)$ we use the following
equations \be  \frac{d \tilde{N}(y)}{d y} - \tilde{C}(y)
\tilde{N}(y)=0~, ~~~ \label{NeqZgc1} \ee where $\tilde{C}(y)$ is
given by the equation \be \label{genericEqNy1}
 \tilde{C}(y)=-\frac{B}{\varrho} ~\frac{\varrho y+\sigma \sqrt{A+B y^2}}{A+B
 y^2}~,
 \ee
and we find in a similar way, as it is described above, that
 \be \label{gcn11}
\tilde{N}(y)=\frac{\tilde{C}_{N}}{\sqrt{A+B y^2}\left|\sqrt{B}
y+\sigma\sqrt{A+ B y^2}\right|^{\frac{ \sqrt{B}}{ \varrho}}} \ee
for $B>0$, and
\be
\tilde{N}(y)=\frac{\tilde{C}_{N}}{\sqrt{A-|B| y^2}}e^{
\frac{\sigma  \sqrt{|B|}}{ \varrho}\tan^{-1}
\left(\frac{\sqrt{|B|} y}{\sqrt{A-| B| y^2}}\right)} \ee for
$B<0$, where $\tilde{C}_{N}$ is an integration constant.

\end{document}